\documentclass[12pt]{article}
\usepackage{epsf,latexsym}
\usepackage[all]{xy}
\usepackage{amsfonts,amssymb} 
\epsfverbosetrue
\newcommand{\de}{\hbox{\rm{d}}}

\newcommand{\pa}{\partial}

\newcommand{\bb}{\begin{eqnarray}}
\newcommand{\ee}{\end{eqnarray}}
\newcommand{\eee}{\nonumber\end{eqnarray}}
\newcommand{\qq}{\quad}

\begin{document}

\font\twelve=cmbx10 at 13pt
\font\eightrm=cmr8

\thispagestyle{empty}

\begin{center}
${}$
\vspace{3cm}

{\Large\textbf{Strong lensing in the Einstein-Straus solution}} \\

\vspace{2cm}

{\large Thomas Sch\"ucker\footnote{also at Universit\'e de Provence, Marseille,
France, thomas.schucker@gmail.com } (CPT\footnote{Centre de Physique
Th\'eorique\\\indent${}$\qq\qq CNRS--Luminy, Case
907\\\indent${}$\qq\qq 13288 Marseille Cedex 9,
France\\\indent${}$\qq
Unit\'e Mixte de Recherche (UMR 6207) du CNRS et des Universit\'es
Aix--Marseille 1 et 2\\
\indent${}$\qq et Sud Toulon--Var, Laboratoire affili\'e \`a la
FRUMAM (FR 2291)})}

\vspace{2cm}

\hfill{\em To the memory of J\"urgen Ehlers}
\vskip 2cm

{\large\textbf{Abstract}}
\end{center}
We analyse strong lensing in the Einstein-Straus solution with positive cosmological constant.
Our result confirms Rindler \& Ishak's finding that a positive cosmological constant decreases the bending of light by an isolated spherical mass. In agreement with an analysis by Ishak et al., this decrease is found to be attenuated by a homogeneous mass distribution added around the spherical mass and by a recession of the observer.
For concreteness we compare the theory to the light deflection of the lensed quasar SDSS J1004+4112.

\vspace{2cm}

\vskip 1truecm

\noindent CPT-P070-2008\\
\noindent 0807.0380
\vspace{1cm}

${}$

\section{Introduction}

In September last year Rindler \& Ishak \cite{ri} corrected the general believe that the deflection angle of light passing near an isolated, static, spherically symmetric mass is independent of the cosmological constant. In their analysis the source emitting the light and the observer were supposed at rest with respect to the central mass and the masses of source and observer were neglected. Two subsequent papers \cite{se,lens} confirmed Rindler \& Ishak's result. Khriplovich \& Pomeransky \cite{kh} pointed out that, if the earth is taken comoving with respect to the exponentially expanding de Sitter space, then the effect of the cosmological constant on the deflection cancels. Park \cite{pa} re-did their analysis with McVittie's solution and finds the same cancelation for the exponentially expanding de Sitter space.

The aim of this paper is to calculate the bending of light by a spherically symmetric mass, which is taken to be a cluster of galaxies, without the two mentioned simplifications: $(i)$ the observer is allowed to move with respect to the cluster, $(ii)$ the masses of the other clusters are included in the form of a homogeneous, isotropic dust. The observer is taken comoving with respect to the dust. This situation is described by the Einstein-Straus solution \cite{es,sch} that matches the Kottler (or Schwarzschild-de Sitter solution) at the inside of the Sch\"ucking radius with a Friedmann solution at the outside. The first motivation of this solution was to explain why the cosmic expansion does not affect small length scales like in solar systems and atoms. Let us note that the Einstein-Straus solution is as unstable as Friedmann's solutions \cite{kr}. This is the very instability that produces structure formation. Ishak et al. \cite{ir} have already used the Einstein-Straus solution in the context of light bending. They find that the dust partially screens the effect of the cosmological constant. Qualitatively this screening is easy to understand: The cosmological constant induces a repulsive force between the isolated cluster and the photon. This force increases with the distance between cluster and photon. Adding more clusters adds more repulsion. But the net force outside the Sch\"ucking radius vanishes due to the high symmetry of the dust. The present calculation will make this screening quantitative. It will show furthermore that the attractive force between cluster and photon, which is due to the central mass and which decreases with distance, is subject to sizable anti-screening. An important part of this anti-screening will turn out to be of purely kinematical origin, coming from the velocity of the observer.

For numerical convenience, we use the following units: length is measured in astro\-meters (am), time in astroseconds (as) and mass in astrograms (ag),
\bb {\rm am}&=&1.30\cdot 10^{26}\ {\rm m}\,=
4221\ {\rm Mpc},\qq\qq\qq
{\rm as}=4.34\cdot 10^{17}\ {\rm s}\, =13.8\ {\rm Gyr},\nonumber\\
{\rm ag}&=&6.99\cdot 10^{51}\ {\rm kg}\,=\, 3.52\cdot 10^{21}\ M_\odot.\ee
In these units, we have  $c=1\ {\rm am\,as^{-1}}$, $8\pi G=1\ {\rm am^3as^{-2}ag^{-1}}$, $H_0=1\ {\rm as^{-1}}$. For completeness we record Planck's constant, which we do not use, $\hbar=3.86\cdot 10^{-121}\ {\rm am^2as^{-1}ag}.$ We will consider
spatially flat universes where we may set the scale factor today $a_0=1$ am. 

\section{Bending of light in Kottler's solution}

Before we take up the Einstein-Straus solution, we review strong lensing in Kottler's solution,
\bb \de \tau^2=B\,\de T^2-\,\frac{1}{B}\,  \de r^2 -r^2 (\sin^2\theta \,\de\varphi ^2+\de
\theta  ^2) ,\qq
B:=1-\,\frac{2GM}{r}\,-\,\frac{\Lambda }{3}\, r^2,\ee
see figure 1, and include a radial velocity of the observer.

\begin{center}
\begin{tabular}{c}

\xy
(0,0)*{}="L";
(-40,0)*{}="T";
(45,-5)*{}="S";
(-40,0)*{\bullet};
(45,-5)*{\bullet};
(49.5,-6)*{S};
(-43,3)*{E};
(-3,-2.5)*{L};
(0,0)*{\bullet};
{\ar (0,0)*{}; (55,0)*{}}; 
"L"; "T" **\dir{-}; 
(-33,-6)*{\alpha };
(-32.5,-1.6)*{\bf\cdot};
(-28,6)*{\alpha '};
(-28.5,1.3)*{\bf\cdot};
(-14,-2)*{r_E};
(23,-5.3)*{r_S};
(11,-0.66)*{\bf\cdot};
(11,-3.8)*{-\varphi _S};
(60,0)*{x};
(0,-19)*{};
"S"; "L" **\dir{-}; 
"T"; "S" **\crv{(0,11)};
"T"; "S" **\crv{(0,-18)};
\endxy
\end{tabular}\linebreak\nopagebreak
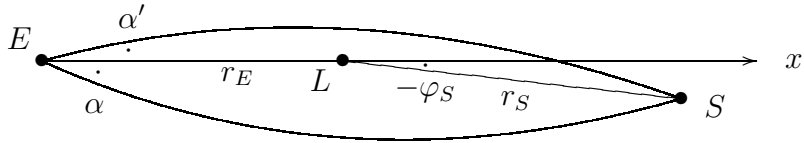
{Figure 1: Two light rays are emitted from the source $S$ and bent by an isolated spherical mass, the lens $L$. They are observed on Earth $E$ under angles $\alpha $ and $\alpha '$.}
\end{center}

In Kottler's solution the geodesics can be integrated analytically to first order in the ratio Schwarzschild radius divided by peri-lens. We are interested in relating the physical observables of strong lensing, the two angles, $\alpha $ and $\alpha '$, between the images and the lens, the redshifts  $z_L$ of the lens and $z_S$ of the source and the mass $M$ of the lens. To be concrete we consider the lensed quasar SDSS J1004+4112 where \cite{in,ot}
\bb \alpha =10''\,\pm\,10\% ,&
 z_L=0.68\ ,&M=5\cdot10^{13}M_\odot \,
 =1.4\cdot10^{-8}\, {\rm ag}\,\pm\,20\%
\\
\alpha ' =\ 5''\,\pm\,10\% ,& z_S=1.734.&
\ee
For this system, the above ratio is of order $10^{-5}$ and second order terms can indeed safely be neglected.

 We use the spatially flat $\Lambda CDM$ model with $\Lambda = 0.77\cdot 3\ {\rm am}^{-2}\ \pm 20\%$ to convert red-shifts into angular distances with respect to the Earth, which we denote by $d_L$ and $d_S$ respectively. Then we obtain the coordinate distances \cite{lens},
\bb d_L=r_E,\qq d_S=\frac{r_E+r_S}{\sqrt{1-\Lambda r_S^2/3}},
\ee
and with the coordinate angle,
\bb \gamma :={\rm arctan}\left[ r_E\left|\frac{\de\varphi }{\de r} \right|_E\right] ,\ee
we get the polar angle of the source,
\bb \varphi '_S \sim \gamma '\left( 1+\,\frac{r_E}{r_S} \right) -\, \frac{4GM}{\gamma' r_E}.\ee 
Notice that this coordinate angle does not depend on the cosmological constant, which however re-enters through
the relation between coordinate angles, $\gamma $, $\gamma '$ and physical angles $\alpha $, 
$\alpha' $:
\bb \tan \alpha \sim\sqrt{1- \Lambda r_E ^2/3}\tan \gamma ,\ee
for an observer at rest with respect to the lens.
From $\varphi _S=\varphi '_S$ we deduce:
\bb \frac{r_E}{r_S} \,\sim\,\frac{4GM}{\alpha \alpha 'r_E}\,(1-\Lambda r_E^2/3)-1\ee
and the mass of the cluster \cite{lat}, see table 1.

\begin{table}[h]
\begin{center}  
\begin{tabular}{|c||c|c|c|c|c|c|c|c|c|c|c|}
\hline
$\Lambda \pm 20\%$&$\pm 0$&$+$&$-$&$ +$&$-$&$+$&$-$&$+$&$-$&+&$-$ \\ 
\hline
$\alpha \pm 10\%$&$\pm 0$&$\pm 0$&$\pm 0$& +&+&$+$&$+$&$-$&$-$&$-$&$-$
                   \\ 
                   \hline
$\alpha' \pm 10\%$&$\pm 0$&$\pm 0$&$\pm 0$& +&+&$-$&$-$&$+$&$+$&$-$&$-$
                   \\ 
\hline\hline
$-\varphi _S\ ['']$&13.0 & 13.6&12.6&15.0&13.9&17.7&16.4&9.5&8.8&12.2&
11.3  \\ 
\hline
$M\  [10^{13}M_\odot]$&4.7 & 5.8& 4.0&7.0&4.8&5.7&$4.0$&$5.7$&$ 4.0$ &4.7&3.2   \\ 
\hline
\end{tabular} 
\end{center}
\caption{ Fitting the cluster mass in Kottler's solution, earth at rest: The coordinate polar angle $\varphi _S$ between Earth and source  and the central mass $M$ are calculated as functions of the cosmological constant and of the measured angles $\alpha $ and $\alpha '$. `$\pm 0$' stands for the central value, '$+$' for the upper and `$-$' for the lower experimental limit.}
\end{table} 

We now want to take into account the velocity $v_E$, that we suppose radially outward. Our task is to recalculate the relation between coordinate angle and physical angle, the latter being measured in nanoseconds over nanoseconds. Consider figure 2 in the $(r,\varphi )$ plane, $\theta =\pi /2$.

\begin{center}
\begin{tabular}{c}

\xy
(0,-9)*{};
(0,15)*{};
(0,0)*{}="L";
(-90,0)*{}="T";
(-30,30)*{}="P";
(-20,0)*{}="A";
(0,0)*{\bullet};
(-90,0)*{\bullet};
(-95,5)*{E};
(2,-3.5)*{L};
(-93,-4.5)*{(r_E,\pi )};
(-15,-4.5)*{(r_E-\de r,\pi )};
(-50,-4.5)*{(r_E-\de r +2\de r_v,\pi )};
(-5,28)*{(r_E-\de r +\de r_v,\pi-\de\varphi  )};
"L"; "A" **\dir{-}; 
(-24,33); "T" **\dir{-}; 
"A"; "P" **\dir{~}; 
"A"; "T" **\dir{~}; 
"P"; (-40,0) **\dir{~};
(-80,0)*{}; (-80.5,5)*{} **\crv{(-77,2.5)};
(-75,3.5)*{ \gamma };

\endxy

\end{tabular}\linebreak\nopagebreak
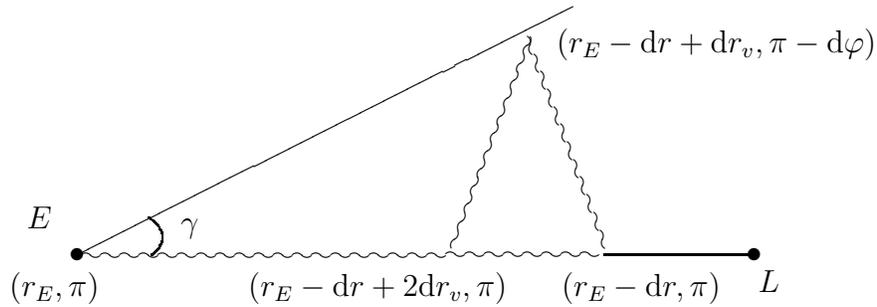
{Figure 2: Relating the coordinate angle $\gamma $ between incoming light rays from the lens and the source to the same angle $\alpha $ as measured in ns/ns. }
\end{center}

 The proper time $\de\tau_r$ it takes a photon to go from $(r_E,\pi )$ to $(r_E-\de r,\pi )$ is computed from
$0=B\,\de T_r^2-(1/B)\,\de r^2,$ with $\de \tau^2_r=B\,\de T_r^2.$ We get $\de\tau_r=B^{-1/2}\de r$. During a lapse $\de\tau_\varphi $ the Earth has moved outwards by $\de r_v=v_E B^{1/2}\de\tau_\varphi = v_EB\,\de T_\varphi $. The proper time $\de \tau_\varphi $ it takes the photon to go from $(r_E-\de r,\pi )$ to $(r_E-\de r+\de r_v,\pi-\de \varphi  )$ is computed from 
$0=B\,\de T_\varphi ^2-(1/B)v_E^2B^2\,\de T_\varphi ^2-r_E^2\,\de\varphi ^2.$ Therefore
$\de\tau_\varphi = B^{1/2}\,\de T_\varphi =(1-v_E^2)^{1/2}r_E\,\de \varphi $. 
Finally
\bb \tan \alpha =\,\frac{\de\tau_\varphi }{\de\tau_r}\, =\,\frac{\sqrt{B}}{\sqrt{1-v_E^2}}\,r_E \,\frac{\de \varphi }{\de r}\, =\,\frac{\sqrt{B}}{\sqrt{1-v_E^2}}\, \tan \gamma .\ee
Imposing again $\varphi _S=\varphi '_S$ we deduce:
\bb \frac{r_E}{r_S} \,\sim\,\frac{4GM}{\alpha \alpha 'r_E}\,\frac{1-\Lambda r_E^2/3}{1-v_E^2}\, -1.\ee
For an Earth comoving with the exponentially expanding de Sitter space, $v_E=\sqrt{\Lambda /3}r_E$, the cosmological constant indeed drops out \cite{kh,pa}. For the more realistic value $v_E=H_0 r_E$ we obtain the values shown in table 2.

\begin{table}[h]
\begin{center}  
\begin{tabular}{|c||c|c|c|c|c|c|c|c|c|c|c|}
\hline
$\Lambda \pm 20\%$&$\pm 0$&$+$&$-$&$ +$&$-$&$+$&$-$&$+$&$-$&+&$-$ \\ 
\hline
$\alpha \pm 10\%$&$\pm 0$&$\pm 0$&$\pm 0$& +&+&$+$&$+$&$-$&$-$&$-$&$-$
                   \\ 
                   \hline
$\alpha' \pm 10\%$&$\pm 0$&$\pm 0$&$\pm 0$& +&+&$-$&$-$&$+$&$+$&$-$&$-$
                   \\ 
\hline\hline
$-\varphi _S\ ['']$&10.5 & 10.4&10.5&11.5&11.6&13.5&13.7&7.3&7.4&9.4&
9.5  \\ 
\hline
$M\  [10^{13}M_\odot]$&3.0 & 3.4& 2.8&4.1&3.4&3.4&$2.7$&$3.4$&$ 2.8$ &2.7&2.3   \\ 
\hline
\end{tabular} 
\end{center}
\caption{ Fitting the cluster mass in Kottler's solution, earth moving with Hubble velocity: The coordinate polar angle $\varphi _S$ between Earth and source  and the central mass $M$ are calculated as functions of the cosmological constant and of the measured angles $\alpha $ and $\alpha '$. `$\pm 0$' stands for the central value, '$+$' for the upper and `$-$' for the lower experimental limit.}
\end{table} 

Taking into account the Hubble velocity of the observer reduces the effect of the cosmological constant on the bending of light: a 20 \% increase of $\Lambda $ decreases the cluster mass by 20 \% for the observer at rest, by only 10 \% for the comoving observer. Consequently the mass estimate of 
$M=4.7^{+2.3}_{-1.5} \cdot 10^{13}M_\odot$ for Kottler's solution with the Earth at rest, which is nicely compatible with the observed value of $M=5.0^{+1.0}_{ -1.0} \cdot 10^{13}M_\odot$ thanks to the positive cosmological constant, is brought down by the Hubble velocity of the observer to 
$M=3.0^{+1.1}_{ -0.7} \cdot 10^{13}M_\odot$,  now only marginally compatible with observation. Naturally we would like to include the effect of the other masses in the universe on the bending of light.

\section{The Einstein-Straus solution with a cosmological constant}

In this section we streamline Sch\"ucking's proof \cite{sch} of the Einstein-Straus solution \cite{es} in its form generalized by Balbinot, Bergamini \& Comastri \cite{bal} to include a cosmological constant. We only consider the case of spatially flat universes. But we add to the results in the above references the Jacobian of the transformation passing between the Friedmann and the Schwarzschild coordinates, which we use in the next section to compute the geodesics of photons.\\[2mm]
{\bf Statement of the result:} We write the Kottler  metric as 
\bb \de \tau^2=B\,\de T^2-\,\frac{1}{B}\,  \de r^2 -r^2 \de
\Omega  ^2 ,\qq
B:=1-\,\frac{2GM}{r}\,-\,\frac{\Lambda }{3}\, r^2,\ee
and the Friedmann metric as
\bb \de \tau^2=\de t^2-a^2[\,\de \chi  ^2+\chi  ^2 \de
\Omega  ^2 ],\qq
\frac{\de a}{\de t}\,=\,\sqrt{ \frac{A}{a}\, +\,\frac{\Lambda }{3}\, a^2},\ee
\vspace {-6mm}
\bb A:={\textstyle\frac{1}{3}}\, 
\rho _{\rm dust\, 0}\,a_0^3.\ee
We suppose that the scale factor $a(t)$ is strictly monotonic. 
Both solutions are glued together at the {\it constant} Sch\"ucking radius $\chi  _{\rm Sch\ddot u}$:
\bb r_{\rm Sch\ddot u}(T):=a(t)\chi  _{\rm Sch\ddot u},\qq r\le  r_{\rm Sch\ddot u},\qq \chi  \ge\chi  _{\rm Sch\ddot u}.\ee
The central mass $M$ must be equal to the dust density times the volume of the ball with Sch\"ucking radius $r_{\rm Sch\ddot u}$,
\bb A=\,\frac{2M}{8\pi \chi _{\rm Sch\ddot u} ^3} \,=\,\frac{2GM}{\chi _{\rm Sch\ddot u}^3} \,.\label{ma}\ee
Then at the Sch\"ucking radius,
\bb B(r_{\rm Sch\ddot u})=:B_{\rm Sch\ddot u}=1-\,\frac{A}{a}\, \chi ^2_{\rm Sch\ddot u}-\,\frac{\Lambda }{3}\, a^2\chi ^2_{\rm Sch\ddot u},\ee
where we also define $C _{\rm Sch\ddot u}:=\sqrt{1-B_{\rm Sch\ddot u}}$.
The coordinate transformation $(T,r)\rightarrow (t,\chi )$ at the Sch\"ucking radius is cumbersome to write down, not so its Jacobian,
\bb \left.\,\frac{\pa t}{\pa T} \right| _{\rm Sch\ddot u}=1,&&
\left.\,\frac{\pa t}{\pa r} \right| _{\rm Sch\ddot u}=-
\,\frac{C _{\rm Sch\ddot u}}{B_{\rm Sch\ddot u}}\,, 
\\[1mm]
\left.\,\frac{\pa \chi }{\pa T} \right| _{\rm Sch\ddot u}=-
\,\frac{C _{\rm Sch\ddot u}}{a}\,,
&&
\left.\,\frac{\pa \chi }{\pa r} \right| _{\rm Sch\ddot u}=
\,\frac{1}{aB_{\rm Sch\ddot u}}\,.
\ee
The inverse of the Jacobian is,
\bb \left.\,\frac{\pa T }{\pa t} \right| _{\rm Sch\ddot u}=
\,\frac{1}{B_{\rm Sch\ddot u}}\,,
&&
\left.\,\frac{\pa T }{\pa \chi } \right| _{\rm Sch\ddot u}=
\,\frac{aC _{\rm Sch\ddot u}}{B_{\rm Sch\ddot u}}\,,
\label{jaci1}\\[1mm]
\left.\,\frac{\pa r }{\pa t} \right| _{\rm Sch\ddot u}=C _{\rm Sch\ddot u},
&&
\left.\,\frac{\pa r }{\pa \chi } \right| _{\rm Sch\ddot u}=a.\label{jaci2}\ee
We will also need to compare coordinate times at the Sch\"ucking radius,
\bb \left.\,\frac{\de t}{\de T} \right| _{\rm Sch\ddot u}=B_{\rm Sch\ddot u}.\label{times}\ee
\\[2mm]
{\bf Proof:} The scale factor $a(t)$ is supposed monotonic and may therefore serve as time coordinate, $(t,\chi )\rightarrow (a,\chi )$. Then the Friedmann metric reads,
\bb \de \tau^2=\,\frac{\de a^2}{A/a+{\textstyle\frac{1}{3}}\Lambda a^2 }\, -a^2\de \chi  ^2-a^2\chi ^2\de\Omega ^2.\ee
In a next step we want to turn the $a^2\chi ^2$ factor in front of $ \de \Omega ^2$ into $r^2$,
\bb (a,\chi )\rightarrow (b,r),\qq a=:\Phi (b,r),\qq \chi =:r/\Phi (b,r),\ee
with the boundary condition that at the Sch\"ucking radius $\chi  _{\rm Sch\ddot u}$, old and new time coordinates coincide, $a=b=\Phi (b,b\chi  _{\rm Sch\ddot u})$.
Then with $C_1:=\sqrt{A/\Phi +{\textstyle\frac{1}{3}}\Lambda \Phi ^2}$ the metric tensor of the Friedmann solution becomes,
\bb g^{\rm Frie}_{bb}&=&\Phi _b^2\,\left\{\,\frac{1}{C_1^2}\, -\,\frac{r^2}{\Phi ^2}\,\right\} ,\qq
 g^{\rm Frie}_{rr}\ =\ -\left[ 1-\,\frac{r}{\Phi }\,\Phi _r\right] ^2+
\,\frac{\Phi _r^2}{C_1^2}\,, \\
g^{\rm Frie}_{br}&=&\Phi _b\,\left\{\,\frac{\Phi _r}{C_1^2}\,  
+\,\frac{r}{\Phi }\, \left[ 1-\,\frac{r}{\Phi }\,\Phi _r\right] \right\} .
\ee
We do not want a mixed term, $g^{\rm Frie}_{br}=0$, which is equivalent to,
\bb \Phi _r=-\,\frac{r}{\Phi }\,\frac{C_1^2}{B_1}\,,\qq
B_1:=1-\,\frac{Ar^2}{\Phi ^3}\,  -\,\frac{\Lambda }{3}\, r^2.\ee
 For every fixed $b$, this differential equation admits one local solution satisfying the boundary condition.
 We can simplify, 
 \bb g^{\rm Frie}_{bb}=\Phi _b^2\,\frac{B_1}{C_1^2}\, 
  ,\qq
 g^{\rm Frie}_{rr}= \,\frac{-1}{B_1}\, .\ee
Differentiating the boundary condition with respect to $b$, we have:
\bb \left.\Phi _b\right|_{\rm Sch\ddot u}:=\Phi_b (b,b\chi _{\rm Sch\ddot u})=
1-\chi _{\rm Sch\ddot u}\left.\Phi _r\right|_{\rm Sch\ddot u}=1/B_{\rm Sch\ddot u}.
\ee
We now turn to the Kottler solution and change its coordinates:
\bb (T,r)\rightarrow (b,r),\qq \,\frac{\de T}{\de b}\, 
=\Psi (b).\label{still}\ee
This coordinate transformation still allows us the choice of one initial condition, which we will use later. In the new coordinates, the metric tensor of the Kottler solution is,
\bb g^{\rm Kott}_{bb}=\Psi ^2 B,\qq 
g^{\rm Kott}_{rr}=-1/B.\ee
It is in these coordinates, $(b,r,\theta ,\varphi )$, that we join together Friedmann's and Kottler's metric tensors continuously at the Sch\"ucking radius and for all times:
\bb \left.g^{\rm Frie}_{bb}\right|_{\rm Sch\ddot u}=
\left.g^{\rm Kott}_{bb}\right|_{\rm Sch\ddot u},\qq
\left.g^{\rm Frie}_{rr}\right|_{\rm Sch\ddot u}=
\left.g^{\rm Kott}_{rr}\right|_{\rm Sch\ddot u}.\ee
At this point we need the relation (\ref{ma}) between  Friedmann's dust density and the central mass $M$. This relation implies $\left.B_1\right|_{\rm Sch\ddot u}=B_{\rm Sch\ddot u}$ and $\left.C_1\right|_{\rm Sch\ddot u}=C_{\rm Sch\ddot u}/\chi _{\rm Sch\ddot u}.$ For the gluing to be continuous we must therefore choose 
\bb\Psi (b)=\,\frac{\chi  _{\rm Sch\ddot u}}{B _{\rm Sch\ddot u}
(b)\,C_{\rm Sch\ddot u}
(b)}\, .\ee
Successive application of the chain rule then gives;
\bb \,\frac{\pa t}{\pa T}\, =\,\frac{\de t}{\de a}\,\frac{\pa \Phi }{\pa b}\,\frac{\de b}{\de T}\, ,&&
\,\frac{\pa t}{\pa r}\, =\,\frac{\de t}{\de a}\,\frac{\pa \Phi }{\pa r}\, ,\\[2mm]
\frac{\pa \chi }{\pa T}\, =\,\frac{\pa \chi  }{\pa b}\,\frac{\de b}{\de T}\, ,\ \qq&&
\,\frac{\pa \chi }{\pa r}\, =\,\frac{1}{\Phi }\,-\,\frac{r}{\Phi ^2}\, \frac{\pa \Phi }{\pa r}\,,\ee
and restricting to the Sch\"ucking radius yields the desired Jacobian.

To compare the Friedmann and Schwarzschild coordinate times $t$ and $T$ at the Sch\"ucking radius, consider the parameterized curve, $T=p,\ r=
\chi  _{\rm Sch\ddot u}b,\ (\theta =\pi /2,\ \varphi =0)$.
Its 4-velocity  is:
\bb  \frac{\de T}{\de p}\, =1,\qq
\frac{\de r}{\de p}\, =\chi  _{\rm Sch\ddot u}\,\left.\frac{\de b}{\de T}\right|_{\rm Sch\ddot u}\frac{\de T}{\de p}\, =B _{\rm Sch\ddot u}\,
C_{\rm Sch\ddot u},\ee
in Schwarzschild coordinates and in Friedmann coordinates:
\bb \frac{\de t}{\de p}& =&\left.\frac{\pa t}{\pa T}\right| _{\rm Sch\ddot u}
\frac{\de T}{\de p}\, +\,\left.\frac{\pa t}{\pa r}\right| _{\rm Sch\ddot u}\frac{\de r}{\de p}\,  =1\cdot 1-
\,\frac{C _{\rm Sch\ddot u}}{B_{\rm Sch\ddot u}}\,B _{\rm Sch\ddot u}=B_{\rm Sch\ddot u},\\[2mm]
\frac{\de \chi }{\de p}& =&\left.\frac{\pa \chi }{\pa T}\right| _{\rm Sch\ddot u}
\frac{\de T}{\de p}\, +\,\left.\frac{\pa \chi }{\pa r}\right| _{\rm Sch\ddot u}\frac{\de r}{\de p}\, =
-\,\frac{C _{\rm Sch\ddot u}}{a}\,\cdot 1+\,\frac{1}{aB_{\rm Sch\ddot u}}\,B _{\rm Sch\ddot u}\,
C_{\rm Sch\ddot u}=0.\ee
Finally we obtain the desired relation: $\de t/\de T=
\de t/\de p\,\cdot\, \de p/\de T=B_{\rm Sch\ddot u}.$

We conclude the proof by pointing out a few typeset errors in reference \cite{bal}: the cosmological constant has the wrong sign in equations (3.15), (4.5) and (4.7). In equation (3.19), $\kappa $ should read $\chi $. In the appendix, the definitions of $S$ and $T$ are missing.

Two remarks are in order: (i) The matching of Kottler's and Friedmann's solutions is possible only if both solutions have the same cosmological constant. (ii) Let us anticipate that the refraction (coordinate) angle $\gamma _F-\gamma _K$, equation (\ref{refraction}), derived from the Jacobian is to first order
\bb \gamma _F-\gamma _K \sim \sqrt{2GM/r_{\rm Sch\ddot u} - {\textstyle\frac{1}{3}}  \Lambda  r^2_{\rm Sch\ddot u} }\,/\cos \gamma _F.\ee
A positive cosmological constant therefore attenuates the refraction.

In the next section we want to interpret the central mass as the mass of a cluster $M\sim 10^{14}\ M_\odot.$ For this interpretation to make sense we must have hierarchies of the following length scales, the Schwarzschild radius $s\sim 10^{-9}$ am, the typical radius of a cluster $r_{\rm cluster}\sim 10^{-3}$ am, the Sch\"ucking radius $r_{\rm Sch\ddot u}\sim 10^{-3}$ am, the typical distance between clusters $D_{\rm cluster}\sim 10^{-3}$ am and the de Sitter radius $r_{\rm dS}\sim 1$ am:
\bb s<r_{\rm cluster}<r_{\rm Sch\ddot u}<D_{\rm cluster}\qq {\rm and}\qq r_{\rm Sch\ddot u}<
r_{\rm dS}.\ee

\section{Integrating the geodesics of light}

The geodesics will be integrated piecewise, see figure 3: in spatially flat Friedmann's solution with cosmological constant $\Lambda = 0.77\cdot 3\ {\rm am}^{-2}$ and dust $\rho _{\rm dust\, 0}=(3-\Lambda )\ {\rm ag/am^3}$ and in Kottler's solution.

\begin{center}
\begin{tabular}{c}

\xy
(0,0)*{}="L";
(-40,0)*{}="T";
(45,-5)*{}="S";
(-40,0)*{\bullet};
(45,-5)*{\bullet};
(49.5,-6)*{S};
(-43,3)*{E};
(-4,-2.5)*{L};
(0,0)*{\bullet};
{\ar (0,0)*{}; (55,0)*{}}; 
"L"; "T" **\dir{-}; 
(-33,-5)*{\alpha };
(-32.5,-0.8)*{\bf\cdot};
(-28,5)*{\alpha '};
(-28.5,0.6)*{\bf\cdot};
(60,0)*{x};
(0,-19)*{};
(0,0)*\cir<5cm>{};
"T"; (-10.7,3.6) **\dir{-};
"S"; (10.9,2.4) **\dir{-};  
(-10.7,3.6); (10.9,2.4) **\crv{(0,3.5)};
"T"; (-9,-6.5) **\dir{-};
"S"; (9,-6.1) **\dir{-}; 
(-9,-6.5); (9,-6.1) **\crv{(-1,-7)}; 
(11,11)*{r_{\rm Sch\ddot u}};
\endxy
\end{tabular}\linebreak\nopagebreak
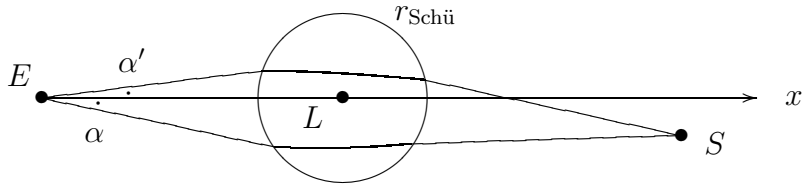
{Figure 3: The two light rays from figure 1 are now bent only inside the Sch\"ucking radius and refracted at the Sch\"ucking radius.}
\end{center}

The geodesics will be pasted together continuously at the Sch\"ucking radius with their first derivatives matched by using the Jacobian computed in the previous section. The scale factor $a(t)$  is computed numerically with a Runge-Kutta method from Friedmann's equation, which in our units reads
\bb 2\,\frac{1}{a}\,a_{tt}\, +
\,\frac{1}{a^2}\left( a_t\right) ^2\,=\,\Lambda ,\ee
with final condition, $a_{t}(0)=a(0)=1.$
 On the other hand, the spatial part of the geodesics is easy to integrate: the photons follow `straight lines' in the polar coordinates $(\chi ,\varphi ,\theta )$. In Kottler's solution the geodesics are integrated manually to first order in the ratio Schwarzschild radius divided by peri-lens.  To be concrete we use again the lensed quasar SDSS J1004+4112, for which the above ratio is of order $10^{-5}$ and second order terms can indeed safely be neglected.

As the physical angles are measured at the arrival, we will integrate backwards in time, i.e. negative $\de t$, $\de T$ and $\de p$, $p$ being the affine parameter. We denote $\de /\de p$ by the over-dot $\dot{}$.

{\bf Step 0} is the integration in Friedmann's solution all the way back to the source without deflection. 

We take the origin, $\chi =0$, at the position of the lens and define the plane containing Earth, lens and source by  
$\theta =\pi /2$. Our final condition at $p=0$ is $t=0,\ \chi =\chi _E,\  \varphi =\pi $ and $\dot t=1,\ \dot\chi =1,\ \dot\varphi =0$. Again we use a Runge-Kutta method to integrate $\de\chi /\de t=1/a$ and we denote the solution by $\tilde\chi (t)$ and its inverse function by $\tilde t(\chi )$. With the definitions of redshift, $1+z=1/a$, and Sch\"ucking radius, 
\bb \chi _{\rm Sch\ddot u}=\left( \frac{3M}{4\pi\rho _{\rm dust\, 0} }\right) ^{1/3},\ee
 we get the values shown in table 3.
 \begin{table}[h]
\begin{center}  
\begin{tabular}{|c||c|c|c|c|c|}
\hline
&Earth&Sch\"ucking radius&lens&Sch\"ucking radius&source\\ 
\hline\hline
$z$&0&&0.68&&1.734\\ 
                   \hline
 $^0t$&0&$-0.4556$&$-0.4566$&$-0.4576$&$-0.7372$                  
 \\
\hline
$\chi $&0.5904&0.0017&0&0.0017&0.5942\\
\hline
\end{tabular} 
\caption{ Passage times of the light ray in astroseconds at the different locations 
and dimensionless comoving coordinate distances of these locations neglecting the bending of the light ray. }
\end{center}
\end{table} 

In {\bf step 1} we compute the trajectory of the photon between the Earth and the Sch\"ucking radius.
Here we need the Christoffel symbols of the Friedmann metric in the plane $\theta =\pi /2$:
\bb {\Gamma ^t}_{\chi \chi }=a_ta,&
{\Gamma ^\chi }_{\varphi \varphi  }=-\chi ,&
 {\Gamma ^\varphi }_{t \varphi  }=a_t/a,\\
 {\Gamma ^t}_{\varphi \varphi  }=\chi ^2a_ta,&
{\Gamma ^\chi }_{t \chi   }=a_t/a,&
 {\Gamma ^\varphi }_{\chi  \varphi  }=1/\chi .\ee
The geodesic reads:
\bb && \ddot t\,+\,a_ta\ \dot \chi ^2\,+\,\chi ^2a_ta\  \dot\varphi ^2\ =\ 0,\\
&&\ddot \chi\, +\,2a_t/a\ \dot t\dot\chi\, -\,\chi\  \dot\varphi 
^2\ =\  0, \\
&& \ddot \varphi\,+\, 2a_t/a\ \dot t\dot\varphi\, +\,2/\chi \ \dot\chi \dot\varphi \ =\ 0.
\ee
To define the final conditions at $p=0$, we use the fact that the coordinate angle arctan$(\chi \dot\varphi /\dot\chi )$ coincides with the physical angle $\alpha '$ measured in nanoseconds/nanoseconds:
\bb t=0,&\chi =\chi _E,&\varphi =\pi ,\\
\dot t=1,& \dot \chi =\cos \alpha ',&
\dot\varphi =\sin\alpha '\,/\chi _E .\ee 
Then the solution of the geodesic equation is unique,
\bb
\dot t=\,\frac{1}{a}\, ,\qq
 \,\frac{\chi _P}{\chi }\, =\,\sin(\varphi -\alpha '),\qq
\dot \varphi \,=\,\frac{\chi _P}{a^2\chi ^2}\,,\ee
where $\chi _P:= \chi _E\sin\alpha '$ is the would-be peri-lens. Therefore the photon crosses the Sch\"ucking sphere in the half-space containing the Earth at the polar-angle 
\bb \varphi _{\rm Sch\ddot u\, E}=\pi -{\rm arcsin}\,\frac{\chi _P}{\chi _{\rm Sch\ddot u}}\, +\alpha ',\ee
 and at the time
\bb t_{\rm Sch\ddot u\, E}=
\tilde t \left( \sqrt{\chi _E^2+\chi _{\rm Sch\ddot u}^2+2\chi _E\chi _{\rm Sch\ddot u}\cos\varphi _{\rm Sch\ddot u\, E}}\right) \ \sim\ ^0t_{\rm Sch\ddot u\, E}.\ee
The difference between $t_{\rm Sch\ddot u\, E}$ and the  non-deflected passage time $^0t_{\rm Sch\ddot u\, E}$ is of second order in $\pi -\varphi _{\rm Sch\ddot u\, E}$.
At this crossing, the 4-velocity of the photon is:
\bb
\dot t_{\rm Sch\ddot u\, E}=\,\frac{1}{a_{\rm Sch\ddot u\, E}}\, ,\qq
 \dot \chi_{\rm Sch\ddot u\, E}=\,-\,\frac{\cos(\varphi_{\rm Sch\ddot u\, E}-\alpha ' )}{a_{\rm Sch\ddot u\, E}^2}\, ,\qq
\dot \varphi_{\rm Sch\ddot u\, E} \,=\,\frac{\chi _P}{a_{\rm Sch\ddot u\, E}^2\chi_{\rm Sch\ddot u} ^2}\,,\ee
with $a_{\rm Sch\ddot u\, E}:=a(t_{\rm Sch\ddot u\, E})$.
Let us call $\gamma _F$, $F$ for Friedmann, the smaller physical angle between the (unoriented) direction of the photon and the  direction towards the lens. We have 
\bb \gamma _F={\rm arctan}\left(\chi_{\rm Sch\ddot u\, E}\, \frac{ \dot \varphi _{\rm Sch\ddot u\, E}}{\dot \chi_{\rm Sch\ddot u\, E}}\right) =\pi -( \varphi _{\rm Sch\ddot u\, E} -\alpha ').\ee

In {\bf step 2} we translate the 4-velocity into the coordinates $T,\ r,\ \varphi $. We now use the free initial condition mentioned after equation (\ref{still}) to set $T_{\rm Sch\ddot u\, E}=t_{\rm Sch\ddot u\, E}$. Using the inverse Jacobian, equation (\ref{jaci2}), we have
\bb \dot r_{\rm Sch\ddot u\, E}=
\,\frac{C_{\rm Sch\ddot u\, E}-\cos(\varphi_{\rm Sch\ddot u\, E}-\alpha ')}{a_{\rm Sch\ddot u\, E}}\,.\ee
Let us call $\gamma _K$, $K$ for Kottler, the smaller {\it coordinate} angle between the (unoriented) direction of the photon and the  direction towards the lens,
\bb \gamma _K:={\rm arctan}\left(r_{\rm Sch\ddot u\, E}\, \frac{ \dot \varphi _{\rm Sch\ddot u\, E}}{\dot r_{\rm Sch\ddot u\, E}}\right)\,=\,
{\rm arctan}\,\frac{\sin \gamma _F}{C_{\rm Sch\ddot u\, E}+\cos\gamma _F}\,.\label {refraction}\ee 
These specify the final conditions for (the spatial part of) the geodesic equation inside the Sch\"ucking sphere. 

In {\bf step 3} we integrate this geodesic equation. To this end we need the Christoffel symbols of the Kottler solution with $\theta =\pi /2$ and denoting $':=\de/\de r$,
\bb {\Gamma ^T}_{Tr}=B'/(2B), &
{\Gamma ^r}_{TT}=BB'/2, &
{\Gamma ^r}_{rr}=-B'/(2B),\\
{\Gamma ^r}_{\varphi \varphi }=-rB, &
{\Gamma ^\varphi }_{r\varphi }=1/r.\ee
The geodesic equations read:
\bb &&\ddot T+B'/B\,\dot T\dot r=0,\\
&&
\ddot r+{\textstyle\frac{1}{2}} BB'\dot T^2-B'/(2B)\,\dot r^2-rB\dot \varphi ^2=0,\\
&&
\ddot \varphi +2r^{-1}\dot r\dot
\varphi =0,
\ee
We immediately get three first integrals:
\bb \dot T=1/B,\qq r^2\dot\varphi =J,\qq
\dot r^2/B+J^2/r^2-1/B=-E.\ee
The last two come from invariance of the metric under rotations and time translations and the integration constants $J$ and $E$ have the meaning of angular momentum and energy per unit of mass. For the photon, $E=0$. Eliminating affine parameter and coordinate time we get:
\bb \frac{\de r}{\de \varphi }=\pm r\sqrt{r^2/J^2-B}.\label{dr}\ee
At the peri-lens $r_P$, $\de r/\de \varphi (r_P)=0$ and therefore $J=r_P B(r_P)^{-1/2}$. Substituting $J$ into equation (\ref{dr}),  the cosmological constant drops out and we have:
\bb\frac{\de \varphi }{\de r}\,= \pm\,
\frac{1}{r\sqrt{r^2/r_P^2-1}}\,\left[ 1-
\,\frac{s}{r}\,-\,\frac{s}{r_P}\,\frac{r}{r+r_P}\right]^{-1/2},\ee
where we have written $s:=2GM$ for the Schwarzschild radius.
From now on we will omit terms of order $(s/r_P)^2$, which in our case are of order $10^{-10}$, and write equalities up to this order with a $\sim$ sign.
In this approximation the peri-lens is
\bb r_P\,\sim\,r_{\rm Sch\ddot u\, E}\sin\gamma _K-{\textstyle\frac{1}{2}} s.\ee
Note that for the upper trajectory, $\de\varphi /\de r$ is positive for $r$ between $r_{\rm Sch\ddot u\, E}$ and $r_P$, negative between
$r_P$ and $r_{\rm Sch\ddot u\, S}$. 
Therefore
\bb \varphi _{\rm Sch\ddot u\, E}-\varphi _{\rm Sch\ddot u\, S}
=\int_{r_P}^{r_{\rm Sch\ddot u\, E}}\left| \frac{\de \varphi }{\de r}\right|\,\de r+\int_{r_P}^{r_{\rm Sch\ddot u\, S}}
\left| \frac{\de \varphi }{\de r}\right|\,\de r.
\ee
Using $\int x^{-1}(x^2-1)^{-1/2}\,\de x=-\arcsin 1/x$, $\int x^{-2}(x^2-1)^{-1/2}\,\de x=(x^2-1)^{1/2}/x,$
$\int (x+1)^{-1}(x^2-1)^{-1/2}\,\de x
=[(x-1)/(x+1)]^{1/2},$ we get to linear order:
\bb \varphi _{\rm Sch\ddot u\, S}&\sim&
\varphi _{\rm Sch\ddot u\, E}-\pi +
{\rm arcsin}\,\frac{r_P}{r _{\rm Sch\ddot u\, E}}\, 
+
{\rm arcsin}\,\frac{r_P}{r _{\rm Sch\ddot u\, S}}\,\nonumber\\[1mm]
\nonumber
&&- {\textstyle\frac{1}{2}} \,\frac{s}{r _{\rm Sch\ddot u\, E}}\,\sqrt{
 \,\frac{r _{\rm Sch\ddot u\, E}^2}{r_P^2}\, -1}
 - {\textstyle\frac{1}{2}} \,\frac{s}{r _{\rm Sch\ddot u\, S}}\,\sqrt{
 \,\frac{r _{\rm Sch\ddot u\, S}^2}{r_P^2}\, -1}\\[1mm]
 &&
 -{\textstyle\frac{1}{2}} \,\frac{s}{r_P}\, 
 \sqrt{\,\frac{r _{\rm Sch\ddot u\, E}-r_P}{r _{\rm Sch\ddot u\, E}+r_P}\, }
  -{\textstyle\frac{1}{2}} \,\frac{s}{r_P}\, 
 \sqrt{\,\frac{r _{\rm Sch\ddot u\, S}-r_P}{r _{\rm Sch\ddot u\, S}+r_P}\, }\,.
\ee 
To proceed we need the coordinate time $T_{\rm Sch\ddot u\, S}$ at which the photon crosses the Sch\"ucking sphere at the source side and its corresponding time $t_{\rm Sch\ddot u\, S}$ in Friedmann's coordinates. Recall that we have defined the time coordinates such that $T_{\rm Sch\ddot u\, E}=t_{\rm Sch\ddot u\, E}$. The time the undeflected photon takes to cross the Sch\"ucking sphere is $^0t_{\rm Sch\ddot u\, E}\,-\,^0t_{\rm Sch\ddot u\, S}=2\cdot10^{-3}$ as. Its time delay due to bending is of the order of 10 years \cite{sz} or $10^{-9}$ as. The difference  $t_{\rm Sch\ddot u\, S}-T_{\rm Sch\ddot u\, S}$ can be estimated with equation (\ref{times}) and an intermediate value theorem:
\bb t_{\rm Sch\ddot u\, S}-T_{\rm Sch\ddot u\, S}
=C^2_{\rm Sch\ddot u\, i}\,(t_{\rm Sch\ddot u\, E}-t_{\rm Sch\ddot u\, S}),\qq C_{\rm Sch\ddot u\, i}^2:=
\,\frac{A}{a(t_i)}\, \chi ^2_{\rm Sch\ddot u}+\,\frac{\Lambda }{3}\, a(t_i)^2\chi ^2_{\rm Sch\ddot u},\ee
with an intermediate value $t_i\in [t_{\rm Sch\ddot u\, S},t_{\rm Sch\ddot u\, E}].$ The function $C_{\rm Sch\ddot u\, i}$ varies slowly, in our example by less than half a per mil, and is small, of the order of $10^{-3}$. We will therefore put $t_{\rm Sch\ddot u\, S}\,=\,^0t_{\rm Sch\ddot u\, S}$ and $r _{\rm Sch\ddot u\, S}=a(^0t_{\rm Sch\ddot u\, S})\chi _{\rm Sch\ddot u}$. 

In {\bf step 4} we translate the four-velocity at $t\,=\,^0t_{\rm Sch\ddot u\, S},\ \chi =\chi _{\rm Sch\ddot u }, \ \varphi  =\varphi  _{\rm Sch\ddot u }$,
\bb \dot T_{\rm Sch\ddot u\, S}=\,\frac{1}{B_{\rm Sch\ddot u\, S}}\, ,\qq
\dot r_{\rm Sch\ddot u\, S}=-\sqrt{1-\,\frac{r_P^2}{r^2_{\rm Sch\ddot u\, S}}\,\frac{B_{\rm Sch\ddot u\, S}}{B_P}\, },\qq
\dot \varphi _{\rm Sch\ddot u\, S}=\,\frac{r_P}{r^2_{\rm Sch\ddot u\, S}\sqrt{B_P}}\, ,\ee
back into Friedmann's solution:
\bb \dot\chi_{\rm Sch\ddot u\, S}\,= \,\frac{-1}{a_{\rm Sch\ddot u\, S}B_{\rm Sch\ddot u\, S}}\, \left( C_{\rm Sch\ddot u\, S}\,+\,\sqrt{1-\,\frac{r_P^2}{r^2_{\rm Sch\ddot u\, S}}\,\frac{B_{\rm Sch\ddot u\, S}}{B_P}\, }\right).\ee
Using the same geometry as in step 1 we get the initial polar-angle of the emitted photon:
\bb \varphi _S=\varphi _{\rm Sch\ddot u\, S}-\gamma _{FS}+
{\rm arcsin}\,\frac{\chi _{\rm Sch\ddot u}\sin\gamma_{FS} }{\chi _S}\,,\qq
\gamma _{FS}:={\rm arctan}\,\frac{-\chi _{\rm Sch\ddot u}\dot\varphi _{\rm Sch\ddot u\, S}}{\dot\chi _{\rm Sch\ddot u\, S}}\, .\ee

\section{Results and conclusion}

First we must point out that the peri-cluster is of the order $r_P\sim 10^{-5}$ am which is very small with respect to the typical radius of a cluster $r_{\rm cluster}\sim 10^{-3}$ am. 

\begin{table}[h]
\begin{center}  
\begin{tabular}{|c||c|c|c|c|c|c|c|c|c|c|c|}
\hline
$\Lambda \pm 20\%$&$\pm 0$&$+$&$-$&$ +$&$-$&$+$&$-$&$+$&$-$&+&$-$ \\ 
\hline
$\alpha \pm 10\%$&$\pm 0$&$\pm 0$&$\pm 0$& +&+&$+$&$+$&$-$&$-$&$-$&$-$
                   \\ 
                   \hline
$\alpha' \pm 10\%$&$\pm 0$&$\pm 0$&$\pm 0$& +&+&$-$&$-$&$+$&$+$&$-$&$-$
                   \\ 
\hline\hline
$-\varphi _S\ ['']$&10.0 & 9.0&10.6&9.9&11.6&11.7&13.7&6.3&7.4&8.1&9.5  \\ 
\hline
$M\  [10^{13}M_\odot]$&1.8 & 1.7& 1.8&2.2&2.2&1.8&$1.8$&$1.8$&$ 1.8$ &1.5&1.5   \\ 
\hline
\end{tabular} 
\end{center}
\caption{ Fitting the cluster mass in Einstein-Straus' solution: The coordinate polar angle $\varphi _S$ between Earth and source  and the central mass $M$ are calculated as functions of the cosmological constant and of the measured angles $\alpha $ and $\alpha '$. `$\pm 0$' stands for the central value, '$+$' for the upper and `$-$' for the lower experimental limit.}
\end{table} 

We compute the two angles $\varphi _S$, one with $\alpha $ and one with $\alpha '$. For the chosen values of the cluster mass $M$ and $\Lambda $, the two angles do not coincide. Even within the error bars for $M,\ \alpha $ and $\alpha '$, there is no value of $\Lambda $ with positive dust-density making the $\varphi _S$s coincide. We therefore keep the experimentally favored cosmological constant $\Lambda = 0.77\cdot 3\ {\rm am}^{-2}\ \pm 20\%$ and fit the mass $M$ in order to achieve coincidence. The results are displayed in table 4.

Taking into account the Hubble velocity of the observer had already reduced the effect of the cosmological constant on the bending of light in Kottler's solution: a 20 \% increase of $\Lambda $ increases the cluster mass by 20 \% for the observer at rest, by only 10 \% for the comoving observer. Now, with realistic velocity and  masses in the universe, an increase of $\Lambda $ by 20 \% only decreases the cluster mass by 5 \%. The dependence on $\Lambda $ comes in step 0 through passage times and comoving distances, in step 2 through the inverse Jacobian and in step 4 through the Jacobian. But at the same time, the central value of the cluster mass has decreased even further, see table 5 and is now incompatible with observation.

\begin{table}[h]
\begin{center}  
\begin{tabular}{|c|c|}
\hline
observation&$M=5.0\matrix{+1.0\cr -1.0} \cdot 10^{13}M_\odot$\\ 
\hline
Kottler, static observer&$M=4.7\matrix{+2.3\cr -1.5} \cdot 10^{13}M_\odot$ 
\\ \hline
Kottler, comoving observer&$M=3.0\matrix{+1.1\cr -0.7} \cdot 10^{13}M_\odot$
\\ \hline
Einstein-Straus&$M=1.8\matrix{+0.4\cr -0.3} \cdot 10^{13}M_\odot$ 
\\ \hline
 
\end{tabular} 
\end{center}
\caption{ Mass estimates for the lensing cluster SDSS J1004+4112 in the three situations: (i) empty universe with a spherical mass $M$ and a static observer, table 1, (ii) empty universe with a spherical mass $M$ and a co-moving observer, table 2, and (iii) dust-filled universe with a spherical mass $M$ and a co-moving observer, table 3 }
\end{table}

There is quite a number of systems where the central mass computed from lensing is up to two times too large compared to the mass inferred from x-rays and it should be interesting to redo the present analysis for those systems. Also the computation of the time delay should be worth to be reconsidered in the Einstein-Straus solution.

\vskip .5cm
\noindent {\bf Acknowledgments:} It is a pleasure to thank Andrzej Sitarz for his warm hospitality at the Jagiellonian University in
Krak\'ow, where part of this work was done within the Transfer of Knowledge Program ``Geometry
in Mathematical Physics''.

\vskip .5cm
\noindent {\bf Note added:}

Between the submission of this paper and its being refereed, five papers appeared, that enrich the controversy about whether or not a cosmological constant modifies the bending of light near an isolated spherical mass. A referee has asked me to comment on these papers.

Sereno \cite{sereno08} again confirms Rindler \& Ishak's findings \cite{ri}.
Gibbons, Warnick \& Werner \cite{gib} re-derive Park's result.
Simpson, Peacock \& Heavens \cite{pea} conclude `that standard results for gravitational lensing in a universe containing $\Lambda $ do not require modification.'
Miraghaei \& Nouri-Zonoz \cite{mira}
confirm Rindler \& Ishak's findings \cite{ri} using a Newtonian limit. 

Finally Ishak, Rindler \& Dossett have written a synthesis \cite{ishak08} comparing the divergent views. They conclude that Khriplovich \& Pomeransky \cite{kh} and Park \cite{pa} have neglected $\Lambda $ terms that are not small enough. I agree with this conclusion, but I would like to go one step further and point out that these  $\Lambda $ terms have an interpretation as physical velocity of the observer's recession from the lens.  In their original paper \cite{ri}, Rindler \& Ishak state explicitly that the observer is at rest with respect to the central mass. The same assumption is made in the earlier literature claiming that the bending of light were independent of $\Lambda $. In my view, this controversy is now settled in favour of Rindler \& Ishak \cite{ri}.

In cosmological applications however, the observer has a non-vanishing velocity. Khri\-plovich \& Pomeransky \cite{kh} and Park \cite{pa} find that for a positive $\Lambda $, there is a particular recession velocity such that the effect of $\Lambda $ on the bending of light cancels. In cosmological situations, this particular velocity is as unrealistic as no velocity. Ishak, Rindler, Dossett, Moldenhauer \& Allison \cite{ir} have taken into account realistic cosmic velocities by the use of an Einstein-Straus solution and find that such realistic velocities  attenuate the effect of the cosmological constant  on the bending of light without however canceling it. This is precisely what the present calculation indicates.


\begin{thebibliography}{10}
\bibitem{ri}
  W.~Rindler and M.~Ishak,
  ``The Contribution of the Cosmological Constant to the Relativistic Bending
  of Light Revisited,''
  Phys.\ Rev.\  D {\bf 76} (2007) 043006
  [arXiv:0709.2948 [astro-ph]].
   \bibitem{se}
  M.~Sereno,
  ``On the influence of the cosmological constant on gravitational lensing in
  small systems,''
  Phys.\ Rev.\  D {\bf 77} (2008) 043004
  [arXiv:0711.1802 [astro-ph]].
    \bibitem{lens}
  T.~Sch\"ucker,
  ``Cosmological constant and lensing,''
  arXiv:0712.1559 [astro-ph], Gen. Rel. Grav. DOI 10.1007/s10714-008-0652-2.
  \bibitem{kh}
  I.~B.~Khriplovich and A.~A.~Pomeransky,
  ``Does Cosmological Term Influence Gravitational Lensing?,''
  arXiv:0801.1764 [gr-qc]. 
  \bibitem{pa}
  M.~Park,
  ``Rigorous Approach to the Gravitational Lensing,''
  arXiv:0804.4331 [astro-ph].
\bibitem{es}
A. Einstein and E. G. Straus,
``The influence of the expansion of space on the gravitation fields surrounding the individual star,''
Rev. Mod. Phys. {\bf 17} (1945) 120, {\bf 18} (1946) 148.
\bibitem{sch}
E. Sch\"ucking,
``Das Schwarzschildsche Linienelement und die Expansion des Weltalls,''
Z.\ Phys.\  {\bf 137} (1954) 595.
\bibitem{kr}
A. Krasi\'nski,
``Inhomogeneous Cosmological Models,''
Cambridge University Press (1997),  page 113.
\bibitem{ir}
  M.~Ishak, W.~Rindler, J.~Dossett, J.~Moldenhauer and C.~Allison,
  ``A New Independent Limit on the Cosmological Constant/Dark Energy from the
  Relativistic Bending of Light by Galaxies and Clusters of Galaxies,'' Mon. Not. R. Astron. Soc. {\bf 388} (2008) 1279
  [arXiv:0710.4726 [astro-ph]].
  \bibitem{in}
  N.~Inada {\it et al.}  [SDSS Collaboration],
  ``A Gravitationally Lensed Quasar with Quadruple Images Separated by 14.62
  Arcseconds,''
  Nature {\bf 426} (2003) 810
  [arXiv:astro-ph/0312427],\\
  M.~Oguri {\it et al.}  [SDSS Collaboration],
  ``Observations and Theoretical Implications of the Large Separation Lensed
  Quasar SDSS J1004+4112,''
  Astrophys.\ J.\  {\bf 605} (2004) 78
  [arXiv:astro-ph/0312429].
  \bibitem{ot}
  N.~Ota {\it et al.},
  ``Chandra Observations of SDSS~J1004+4112: Constraints on the Lensing Cluster
and Anomalous X-Ray Flux Ratios of the Quadruply Imaged Quasar,''
  Astrophys.\ J.\  {\bf 647} (2006) 215
  [arXiv:astro-ph/0601700].
  \bibitem{lat}
  T.~Sch\"ucker,
  ``Strong lensing with positive cosmological constant,''
  arXiv:0805.1630 [astro-ph], Moriond Proceedings `Cosmology 2008'.
\bibitem{bal}
R. Balbinot, R. Bergamini and A. Comastri, ``Solution of the Einstein-Straus problem with a $\Lambda $ term,'' Phys.\ Rev.\  D {\bf 38} (1988) 2415.
  \bibitem{sz}
 T.~Sch\"ucker and N.~Zaimen,
 ``Cosmological constant and time delay,''
   A\&A {\bf 484}  (2008) 103 [arXiv:0801.3776 [astro-ph]].
  \bibitem{sereno08}
  M.~Sereno,
 ``The role of Lambda in the cosmological lens equation,''
  arXiv:0807.5123 [astro-ph].
\bibitem{gib}
  G.~W.~Gibbons, C.~M.~Warnick and M.~C.~Werner,
 ``Light-bending in Schwarzschild-de-Sitter: projective geometry of the
  optical metric,''
  arXiv:0808.3074 [gr-qc].
  \bibitem{pea}
  F.~Simpson, J.~A.~Peacock and A.~F.~Heavens,
  ``On lensing by a cosmological constant,''
  arXiv:0809.1819 [astro-ph].
  \bibitem{mira}
  H.~Miraghaei and M.~Nouri-Zonoz,
 ``Classical tests of general relativity in the Newtonian limit of
  Schwarzschild-de Sitter spacetime,''
  arXiv:0810.2006 [gr-qc].
\bibitem{ishak08}
  M.~Ishak, W.~Rindler and J.~Dossett,
  ``More on Lensing by a Cosmological Constant,''
  arXiv:0810.4956 [astro-ph].

 

\end{thebibliography}
\end{document}